\documentclass[showpacs,prl,reprint,preprintnumbers,amsmath,amssymb,superscriptaddress,nofootinbib]{revtex4-1}

\usepackage{amsmath,setspace,subfigure,amsfonts,latexsym}
\usepackage{amssymb}
\usepackage{color}
\usepackage{epsfig}
\usepackage{color}
\usepackage{hyperref}
\usepackage[compat=1.1.0]{tikz-feynman}
\tikzfeynmanset{
every edge={thick},
}

\definecolor{White}{rgb}{1,1,1}
\definecolor{Red}{rgb}{1,0.1,0}
\definecolor{LightYellow}{rgb}{1,1,.875}
\definecolor{SteelBlue}{rgb}{.273,.508,.703}
\definecolor{navy}{rgb}{0,0,.5}
\definecolor{LightCyan}{rgb}{.875,1,1}
\definecolor{DarkRed}{rgb}{.543,0,0}
\definecolor{HotPink}{rgb}{1,.41,.70}
\definecolor{ForestGreen}{rgb}{.13,.54,.13}
\definecolor{OliveDrab}{rgb}{.42,.55,.14}
\definecolor{MediumBlue}{rgb}{0,0,.80}
\definecolor{RoyalBlue}{rgb}{.25,.41,.88}
\definecolor{DeepSkyBlue}{rgb}{0,.746,1}
\definecolor{Brown}{rgb}{0.545,0.271,0.074}

\def\bec{\begin{center}}
\def\ec{\end{center}}

\def\beq{\begin{equation}}
\def\eeq{\end{equation}}

\newcommand\lsim{\mathrel{\rlap{\lower4pt\hbox{\hskip1pt$\sim$}}
    \raise1pt\hbox{$<$}}}
\newcommand\gsim{\mathrel{\rlap{\lower4pt\hbox{\hskip1pt$\sim$}}
    \raise1pt\hbox{$>$}}}
\def\bea{\begin{eqnarray}}
\def\eea{\end{eqnarray}}
\def\ba{\begin{array}}
\def\ea{\end{array}}

\newcommand{\axino}{{\tilde a}}
\newcommand\unit[1]{\,{\rm #1}}
\newcommand\eV{\unit{eV}}
\newcommand\keV{\unit{keV}}
\newcommand\MeV{\unit{MeV}}
\newcommand\GeV{\unit{GeV}}
\newcommand\TeV{\unit{TeV}}

\begin{document}

\title{Colder Freeze-in Axinos Decaying into Photons}
\author{Kyu Jung Bae} 
\author{Ayuki Kamada} 
\affiliation{Center for Theoretical Physics of the Universe, Institute for Basic Science (IBS), Daejeon 34051, Korea}
\author{Seng Pei Liew}
\affiliation{Physik-Department, Technische Universit\"at M\"unchen, 85748 Garching, Germany}
\author{Keisuke Yanagi}
\affiliation{Department of Physics, University of Tokyo, Bunkyo-ku, Tokyo 113-0033, Japan}

\date{\today}

\preprint{
CTPU-17-24,~ UT-17-24
}

\begin{abstract}
We point out that $7 \keV$ axino dark matter (DM) in the R-parity violating (RPV) supersymmetric (SUSY) Dine-Fischler-Srednicki-Zhitnitsky model can simultaneously reproduce the $3.5 \keV$ X-ray excess, and evade stringent constraints from the Ly-$\alpha$ forest data.
Peccei-Quinn symmetry breaking naturally generates both axino interactions with minimal SUSY standard model particles and RPV interactions.
The RPV interaction introduces an axino-neutrino mixing and provides axino DM as a variant of sterile neutrino DM, whose decay into a monochromatic photon can be detected by X-ray observations.
Axinos, on the other hand, are produced by freeze-in processes of thermal particles in addition to the Dodelson-Widrow mechanism of sterile neutrinos.
The resultant phase space distribution tends to be colder than the Fermi-Dirac distribution.
The inherent entropy production from late-time saxion decay makes axinos even colder.
The linear matter power spectrum satisfies even the latest and strongest constraints from the Ly-$\alpha$ forest data.

\end{abstract}

\maketitle

%\titlepage

{\it Introduction} -- Supersymmetry (SUSY) and Peccei-Quinn (PQ) symmetry are two of the most promising extensions of the standard model (SM).
While SUSY resolves the hierarchy between the electroweak scale and the quantum gravity or grand unification scale~\cite{susy},
PQ symmetry explains why quantum chromodynamics (QCD) preserves $CP$ symmetry accurately~\cite{pq}.
The extensions of the SM with these symmetries introduce natural dark matter (DM) candidates: neutralino and axion, respectively.
While a promising parameter region of axion DM is still under investigation~\cite{admx}, neutralino DM is already tightly constrained by direct and indirect searches~\cite{wimp}.
On the other hand, combining the two symmetric extensions introduces another attractive DM candidate: axino ($\axino$), which is the fermion SUSY partner of QCD axion ($a$)~\cite{Rajagopal:1990yx}.
The mass of axino is generated by SUSY breaking, so naively is of order the gravitino mass.
In some models, however, the axino mass can be of order keV~\cite{axinomass}, where axino is the lightest SUSY particle and thus a warm dark matter (WDM) candidate.

In this letter, we consider the R-parity violating (RPV) SUSY Dine-Fischler-Srednicki-Zhitnitsky (DFSZ) model~\cite{dfsz}.
The SUSY $\mu$-term is generated by PQ symmetry breaking \`a la the Kim-Nilles mechanism~\cite{Kim:1983dt}.
The interaction responsible for the $\mu$-term mediates axino and minimal supersymmetric standard model (MSSM) particles.
The production of axinos follows the {\it freeze-in} nature of feebly interacting massive particles (FIMPs)~\cite{Hall:2009bx}, since the interactions with MSSM particles are apparently renormalizable but are feeble due to the suppression by the PQ symmetry breaking scale, $v_{\rm PQ} \gtrsim 10^{9} \GeV$.
In addition, bilinear R-parity violating (bRPV) terms are generated in a similar way~\cite{axino_neutrino}.
Such interactions trigger an axino-neutrino mixing like sterile neutrino~\cite{Abazajian:2012ys}, and the resultant axino DM decay into a monochromatic photon can be detected by X-ray observations.

In light of the recent evidence of an anomalous $3.5 \keV$ X-ray line in the Andromeda galaxy and galaxy clusters, it is timely to consider models of keV-mass decaying DM.
The line excess in the \texttt{XMM-Newton} and \texttt{Chandra} data was first reported by two independent groups in February 2014~\cite{Boyarsky:2014jta, Bulbul:2014sua}. 
Subsequent studies showed that similar excesses are also found in the Galactic Center~\cite{Boyarsky:2014ska} and in the \texttt{Suzaku} data~\cite{Urban:2014yda}. 
While there are reports of a null detection ({\it e.g.}, in observations of dwarf spheroidal galaxies~\cite{Malyshev:2014xqa}), the decaying DM explanation of the $3.5 \keV$ line excess is yet to be excluded (see~\cite{Iakubovskyi:2015wma} for a thorough review).

Constraints from Ly-$\alpha$ forest data are rather relevant, in general, when decaying $7 \keV$ DM is considered as an origin of the $3.5 \keV$ line excess.
For example, $7 \keV$ sterile neutrino DM from scalar particle decay satisfies the less stringent constraints, $m_{\rm WDM} > 2.0$~\cite{Viel:2005qj} and $3.3 \keV$~\cite{Viel:2013apy}, but is in tension with the recently updated ones, $m_{\rm WDM} > 4.09$~\cite{Baur:2015jsy} and $5.3 \keV$~\cite{Irsic:2017ixq}, as shown by direct comparisons of linear matter power spectra~\cite{sterile_production}.
One may wonder why $m_{\rm WDM} > 5.3 \keV$ disfavors $7 \keV$ DM.
Note that such bounds are derived under the assumption that WDM particles follow the Fermi-Dirac distribution with two spin degrees of freedom, and they reproduce the observed DM mass density by tuning the temperature ($T_{\rm WDM}$) for a given mass ($m_{\rm WDM}$):
\begin{eqnarray}
\Omega_{\rm WDM} h^{2} &=& \left( \frac{m_{\rm WDM}}{94 \eV} \right) \left( \frac{T_{\rm WDM}}{T_{\nu}} \right)^{3} \notag \\
&=& 7.5 \left( \frac{m_{\rm WDM}}{7 \keV} \right) \left( \frac{106.75}{g_{*} (T_{\rm dec})} \right) \,.
\end{eqnarray}
where $T_{\nu}$ is the neutrino temperature, and $g_{*} (T_{\rm dec})$ is the effective massless degrees of freedom when the WDM particles are decoupled.
This relation clearly shows that we need $g_{*} (T_{\rm dec}) \sim 7000$, which implies a large entropy dilution factor, $\Delta \sim 70$, in addition to the full SM degrees of freedom, $g_{\rm SM} = 106.75$, even if the WDM particles decouple before the electroweak phase transition.

The phase space distribution of freeze-in axinos varies depending on its production processes, and thus it is affected by the mass spectrum of MSSM particles involved in freeze-in processes.
We obtain the resultant phase space distribution by integrating the Boltzmann equation, and find that it is typically {\it colder} than the Fermi-Dirac one.%
\footnote{Different realizations of $7 \keV$ axino DM decay were considered in Refs.~\cite{axino_decay}.
Nevertheless, none of them discussed a phase space distribution of axino DM or Ly-$\alpha$ forest constraints.}
Saxion ($s$), which is the scalar partner of axion, also makes axinos DM {\it colder}, since its late-time decay injects a certain amount of entropy to the thermal bath after axino decoupling.
We calculate the resultant linear matter power spectra of freeze-in $7 \keV$ axino DM, and show that they are concordant with the current constraints from the Ly-$\alpha$ forest data.

{\it Model} -- The DFSZ solution to the strong $CP$ problem invokes a coupling between a PQ symmetry breaking field ($X$) and the up- and down-type Higgs doublets $\left( H_{u, d} \right)$.
Its SUSY realization is given by the following superpotential:
\begin{equation}
W_{\rm DFSZ} = \frac{y_{0}}{M_{*}} X^{2} H_{u} H_{d} \,,
\end{equation}
where $y_{0}$ is a dimensionless constant and $M_{*}$ is a cutoff scale.
The PQ charges of $X$, $H_{u}$, and $H_{d}$ are respectively $-1$, $1$, and $1$. 
Once the field $X$ develops its vacuum expectation value (VEV), {\it i.e.}, $X = (v_{\rm PQ} / \sqrt{2}) \exp(A / v_{\rm PQ})$, where $A = (s + i a) / \sqrt{2} + \sqrt{2} \theta \axino + \theta^{2} {\cal F}_A$ is the axion superfield, the $\mu$-term and an axino interaction are generated as
\begin{equation}
W_{\rm DFSZ} = \mu e^{2 A / v_{\rm PQ}} H_{u} H_{d} \simeq \mu \left( 1 + \frac{2A}{v_{\rm PQ}} \right) H_{u} H_{d} \,,
\end{equation}
where $\mu = y_{0} v_{\rm {PQ}}^{2} / (2 M_{*})$.
The approximate equality is valid when one considers the axino interaction.
If $M_{*} \sim 10^{16} \GeV$, $y_{0} \sim 0.1$, and $v_{\rm PQ} \sim 10^{10} \GeV$, one finds $\mu \sim 500 \GeV$.
This is a well-known solution to $\mu$-term generation by the Kim-Nilles mechanism~\cite{Kim:1983dt}.
From this renormalizable interaction, freeze-in production of axinos occurs dominantly when the cosmic temperature ($T$) is of order the mass of the other SUSY particle involved in the process~\cite{Chun:2011zd, Bae:2011jb, Bae:2011iw}. 
The contributions from dimension-five anomaly operators ({\it e.g.}, axino-gluino-gluon) are suppressed~\cite{Bae:2011jb}.

The bRPV term is also generated as~\cite{axino_neutrino}
\begin{equation} 
W_{\rm bRPV} = \frac{y'_{i}}{M_{*}^{2}} X^{3} L_{i} H_{u}
\simeq \mu'_{i} \left( 1 + \frac{3A}{v_{\rm PQ}} \right) L_{i} H_{u},
\end{equation}
If $M_{*} \sim10^{16} \GeV$, $y'_{i} \sim1$, and $v_{\rm PQ} \sim 10^{10} \GeV$, one finds $\mu'_{i} \sim {\rm MeV}$.
This term generates mixing between active neutrinos and axino.
The mixing angle is given by
\begin{equation}
|\theta| \simeq \frac{\mu' v_{u}}{m_{\axino} v_{\rm PQ}} \simeq 10^{-5} \left( \frac{\mu'}{4 \MeV} \right)
\left( \frac{7 \keV}{m_{\axino}} \right) \left( \frac{10^{10} \GeV}{v_{\rm PQ}} \right) \,,
\end{equation}
where $v_{u}$ is the VEV of $H_{u}$ and $m_{\axino}$ is the axino mass.
One finds that the mixing parameter of $\sin^{2} 2 \theta \sim 10^{-10}$ is easily obtained, so axino DM decay can be an origin of the $3.5 \keV$ X-ray line excess like sterile neutrino DM.
From this mixing, axinos are produced by the Dodelson-Widrow mechanism~\cite{Dodelson:1993je}, but they account for only a few \% of the total DM density~\cite{sterile_density}.
Therefore, there must exist a more efficient production mechanism of axinos: {\it freeze-in production} via the $\mu$-term interaction.

\begin{figure}
  \centering
      \includegraphics[width=0.45\textwidth]{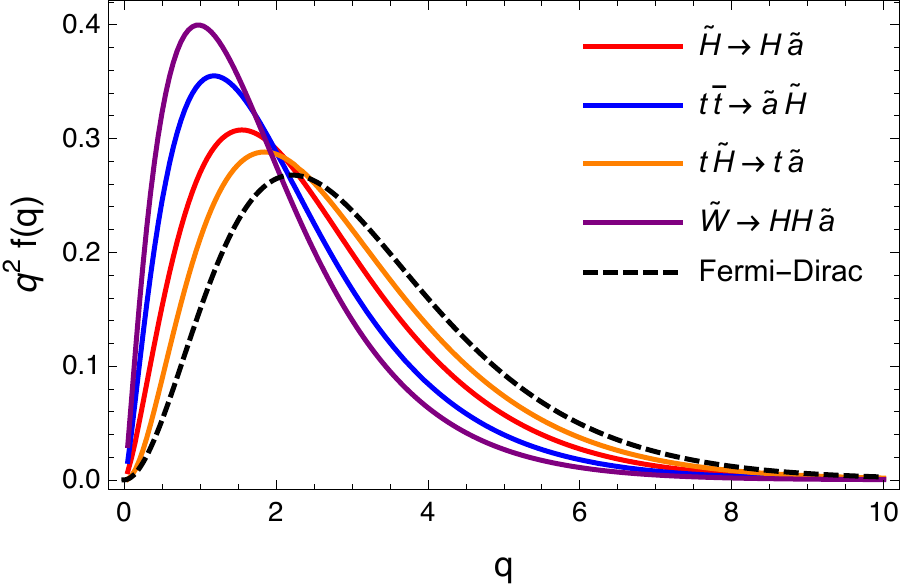}
  \caption{Axino phase space distributions from respective production processes. 
  The red, blue, and yellow solid lines show $q^2 f(q)$ respectively from Higgsino 2-body decay and s- and t-channel scatterings, while the purple solid line shows that from wino 3-body decay. 
  For comparison, the Fermi-Dirac distribution is shown by the dashed line. 
  Each distribution is normalized such that $\int dq q^2 f(q) = 1$.}
  \label{fig:dist-process}
\end{figure}
{\it Freeze-in Production} -- The production of axino is governed by the following Boltzmann equation:
\begin{align}
  \label{eq:boltzmann}
  \frac{df_{\axino}(t, p)}{dt}
  = \frac{\partial f_{\axino}(t, p)}{\partial t} - \frac{1}{R(t)} \frac{dR(t)}{dt} p \frac{\partial f_{\axino}(t, p)}{\partial p}
  = \frac{1}{E} C(t, p) \,,
\end{align}
where $f_{\axino}(t, p)$ is the axino phase space distribution as a function of the cosmic time ($t$) and the axino momentum ($p$), $R(t)$ is the cosmic scale factor, $E$ is the axino energy, and $C(t, p)$ is the collision term.
Due to feeble interactions of axino, one can safely neglect $f_{\axino}$ in the collision term.
Then by integrating the both sides from $t=t_{i}$ to $t=t_{f}$, one finds
\begin{equation}
f_{\axino}(t_{f}, p) = \int^{t_{f}}_{t_{i}} dt\frac{1}{E}
C\left(t,\frac{R(t_{f})}{R(t)}p\right).
\end{equation}
Once one collects all the relevant contributions to the collision term, it is easy to obtain the axino phase space distribution.
We do not provide details here, but refer readers to Ref.~\cite{bkly}.

For the freeze-in production of axinos, the contributions of 2-body and 3-body decays, and s- and t-channel scatterings are taken into account.
In Fig.~\ref{fig:dist-process}, phase space distributions, in form of $q^{2} f_{\axino}(q)$ ($q = p_{\axino} / T_{\axino}$), are shown for Higgsino 2-body decay ($\widetilde{H} \to H + \axino$), s-channel scattering ($t + {\bar t} \to \widetilde{H} + \axino$), t-channel scattering ($\widetilde{H} + t \to \axino + t$), and wino 3-body decay ($\widetilde{W} \to H + H + \axino$).%
\footnote{In Fig.~\ref{fig:dist-process}, we take tops ($t$ and ${\bar t}$) and Higgses to be massless, while introducing the thermal mass of intermediate Higgs in t-channel scattering. 
In the realistic analysis with the benchmark point below, however, we take into account the Higgs soft masses while tops are still massless.}
Here we define the axino temperature by $T_{\axino} = (g_{*}(T) / g_{*}(T_{\rm th}))^{1/3} T$, where $T_{\rm th}$ is set to the mass of the other SUSY particle involved in the freeze-in process.
While all the freeze-in processes shown in Fig.~\ref{fig:dist-process} have a {\it colder} phase space distribution than the Fermi-Dirac distribution, a 3-body decay case has the coldest distribution.
The reason is that 3-body decay leads to a smaller kinetic energy of the final-state axino than the other processes at a given temperature.
However, when one considers a realistic example, such a 3-body decay rarely dominates over other processes, so the resulting axino phase space distribution follows those of 2-body decay or s- and t-channel scatterings.

For a realistic analysis, we consider a benchmark point where the Higgsino-like neutralino is the next-to-lightest SUSY particle (NLSP).
The mass spectrum is shown in Table~\ref{tab:spectra}.
\begin{table}
  \centering
  \begin{tabular}{cccc}\hline
    ~Higgs VEV ratio~ & ~$\tan\beta$~ & ~20~ \\
    ~$\mu$-term~ & ~$\mu$~ & ~$500 \GeV$~  \\
    ~wino mass~ & ~$M_{2}$~ & ~$10000 \GeV$~  \\
    ~$CP$-odd Higgs mass~ & ~$m_{A}$~ & ~$10000 \GeV$~  \\
    ~stop masses~ & ~$m_{\widetilde{Q}_{3}} = m_{\tilde{t}^{c}}$~ & ~$6500 \GeV$~  \\
    ~SM-like Higgs mass~ &~ $m_{h}^{\text{SM-like}}$~ & ~$125 \GeV$~  \\
      \hline
      ~$H_{u}$ soft mass~&~$m_{H_{u}} (Q = m_{\tilde{t}^{c}})$~ & ~$956 \GeV$~  \\
      ~$H_{d}$ soft mass~&~$m_{H_{d}} (Q = m_{\tilde{t}^{c}})$~ & ~$9940 \GeV$~ \\
      \hline
  \end{tabular}
  \caption{MSSM parameters of the benchmark point with Higgsino NLSP is shown.
  The SM-like Higgs mass and soft masses at $Q = m_{\tilde{t}^{c}}$ are calculated by \texttt{SUSY-HIT v1.5a}~\cite{Djouadi:2006bz}. 
  The masses of all the other SUSY particles are taken to be $10 \TeV$.}
  \label{tab:spectra}
\end{table}
In this benchmark scenario, the dominant process is Higgs decay into Higgsino and axino, while Higgsino 3-body decay and s- and t-channel scatterings also contribute.
\begin{figure}
  \centering
   \includegraphics[width=0.45\textwidth]{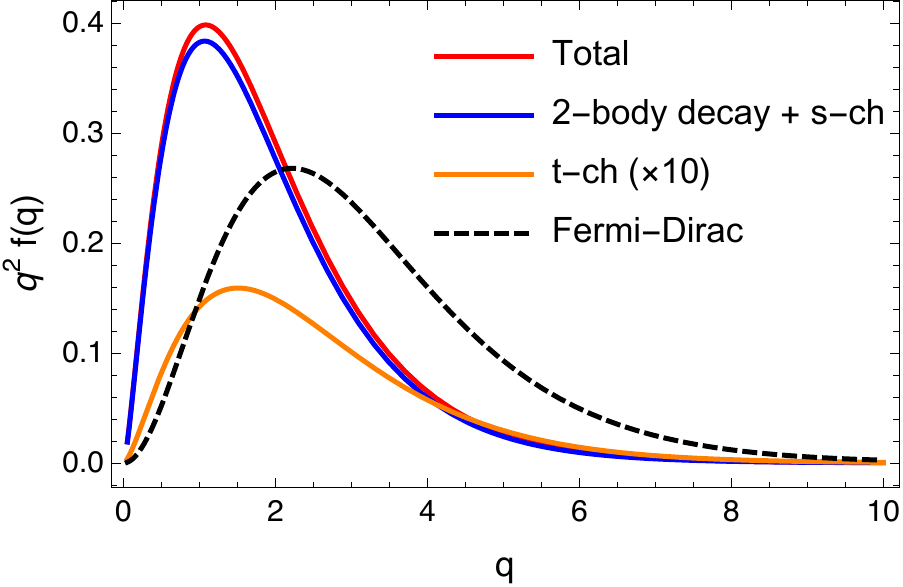}  
  \caption{Axino phase space distributions for the benchmark point. 
  The red solid line shows the total axino phase space distribution normalized such that $\int dq q^2 f(q) = 1$. 
  The blue solid line is sum of the contributions from Higgs 2-body decay and Higgsino s-channel scattering, and the yellow solid line is the contribution from Higgsino t-channel scattering (multiplied by $10$ for visualization). 
  The normalized Fermi-Dirac distribution is shown by the dashed line.}
  \label{fig:dist-real}
\end{figure}
Figure~\ref{fig:dist-real} shows the resultant axino phase space distribution accompanied by the contributions of the respective processes.%
\footnote{We add the s-channel scattering contribution to that of the 2-body decay.
This is because we define the s-channel scattering contribution by subtracting the Higgs pole from the matrix element to avoid the double counting of the 2-body decay.
See Ref.~\cite{bkly} for details.} 
It is clearly shown that the freeze-in production of axinos leads to a colder phase space distribution than the Fermi-Dirac one.

{\it Ly-$\alpha$ Constraints} -- In order to examine whether $7 \keV$ freeze-in axino DM with the phase space distribution obtained above is concordant with the constraints from the Ly-$\alpha$ forest data, we calculate linear matter power spectra by using a Boltzmann solver, {\tt CLASS}~\cite{class}.
We define the squared transfer function by the ratio of the WDM linear matter power spectrum to the cold dark matter one, which is denoted by ${\cal T}^2(k)$ as a function of the wave number, $k$.
\begin{figure}
  \centering
    \includegraphics[width=0.45\textwidth]{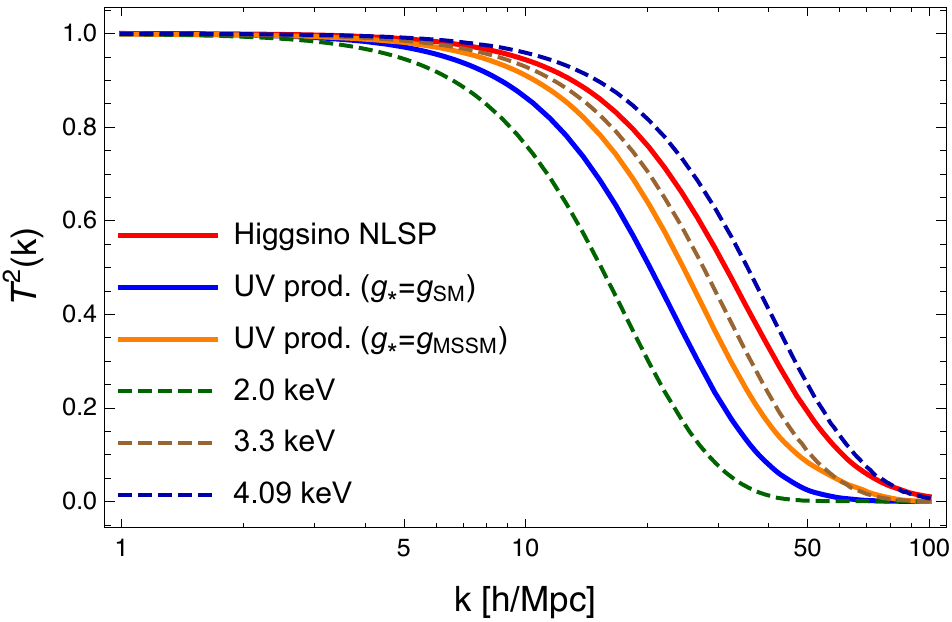}
  \caption{Squared transfer functions for 7 keV axino DM and Ly-$\alpha$ forest constraints. 
  The red solid line shows ${\cal T}^{2} (k)$ in the benchmark point, and the blue (yellow) solid line shows ${\cal T}^{2} (k)$ for axino DM from UV production with $g_{*} = g_{\rm SM} = 106.75$ ($g_{*} = g_{\rm MSSM} = 226.75$). 
  The green, brown, and blue dashed lines respectively show ${\cal T}^{2} (k)$ for $m_{\rm WDM} = 2.0, 3.3$, and $4.09 \keV$.}
  \label{fig:power}
\end{figure}
Figure~\ref{fig:power} compares ${\cal T}^2(k)$ in the benchmark point with those for the Ly-$\alpha$ forest lower bounds of $m_{\rm WDM} = 2.0, 3.3$, and $4.09 \keV$.
For comparison, we also show ${\cal T}^2(k)$ for $7 \keV$ axino DM from UV production via non-renormalizable operators (more specifically, $\widetilde{W} + H \to H + \axino$), where the produced axinos follow the Fermi-Dirac distribution.%
\footnote{In this case, the axino phase space distribution is slightly different from the Fermi-Dirac one, since axinos are not thermalized~\cite{bkly}.}
It is clearly shown that $7 \keV$ axino DM from UV production is disfavored by the Ly-$\alpha$ forest data, when one incorporates the $m_{\rm WDM} > 3.3 \keV$ or stronger constraint.
On the contrary, $7 \keV$ axino DM from freeze-in production in our benchmark scenario shows larger ${\cal T}^2(k)$ so that it is allowed even by the constraint of $m_{\rm WDM} > 3.3 \keV$.
It is, however, still in tension with the stronger constraint, $m_{\rm WDM} > 4.09 \keV$.

In this regard, one can conclude that a certain amount of entropy production is still necessary, when the stronger Ly-$\alpha$ forest constraints, $m_{\rm WDM} > 4.09$ and $5.3 \keV$, are taken into account.
\begin{figure}
  \centering
  \includegraphics[width=0.45\textwidth]{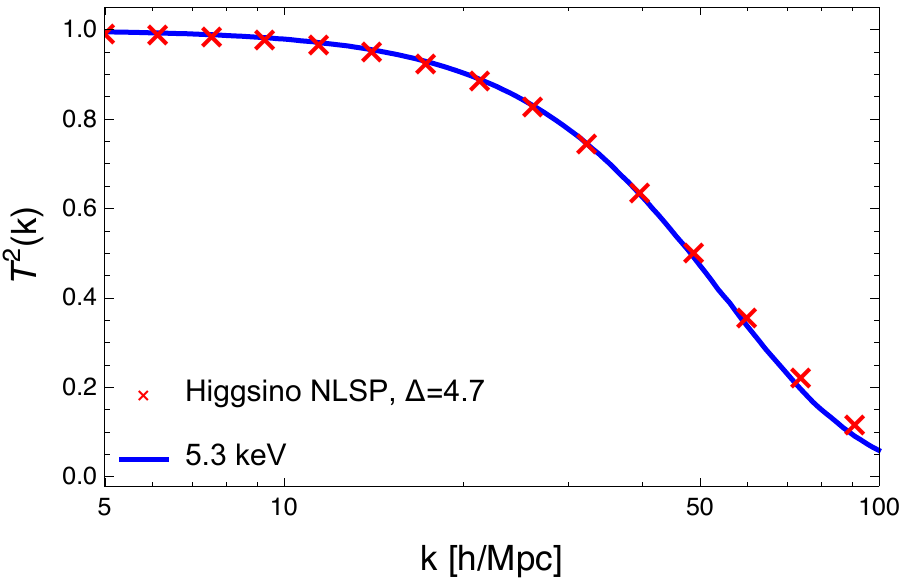}
  \caption{Squared transfer functions with the entropy production from late-time saxion decay. 
  Red crossed points show ${\cal T}^{2} (k)$ for the benchmark points with $\Delta = 4.7$. 
  Blue solid line shows ${\cal T}^{2} (k)$ for $m_{\rm WDM} = 5.3 \keV$ corresponding to the most stringent lower bound from the Ly-$\alpha$ forest data. 
  }
  \label{fig:T2k-entropy}
\end{figure}
In Fig.~\ref{fig:T2k-entropy}, we find that $7 \keV$ axino DM with $\Delta = 4.7$ fits the strongest lower bound from the Ly-$\alpha$ forest data, $m_{\rm WDM} = 5.3 \keV$, very well.%
\footnote{We can infer the entropy dilution factor by comparing the second moments of the phase space distribution of $7 \keV$ axino DM and of the Fermi-Dirac one with $m_{\rm WDM} = 5.3 \keV$~\cite{bkly, Kamada:2013sh}.}
It means that we need only a mild entropy dilution factor, $\Delta > 4.7$, to evade the Ly-$\alpha$ forest constraints.

In the SUSY DFSZ model, such an entropy dilution factor is easily achieved by late-time saxion decay.
Saxions are abundantly produced in the form of coherent oscillation; the yield is given by
\begin{equation}
  Y_{s}^{\rm CO} \simeq 1.9 \times 10^{-6} \left( \frac{\rm GeV}{m_{s}} \right) \left( \frac{{\rm min}[T_{R},T_{s}]}{10^7 \GeV} \right) \left( \frac{s_{0}}{10^{12} \GeV} \right)^{2} \,,
\end{equation}
where $m_{s}$ is the saxion mass, $T_{R}$ is the reheat temperature, $s_{0}$ is the saxion initial amplitude, and $T_{s}$ is determined by $(3 / R) \, (dR / dt) |_{T = T_{s}} = m_{s}$.
Such saxions dominates the energy density of the Universe at the temperature,
\begin{equation}
T_{e}^{s} \simeq 2.5 \times 10^{2} \GeV
\left( \frac{{\rm min}[T_{R},T_{s}]}{10^{7} \GeV} \right) \left( \frac{s_{0}}{10^{16} \GeV} \right)^{2} \,.
\end{equation}
For $\Delta = 4.7$, it is required that saxion decay occurs at $T = T_{D}^{s} \simeq 53 \GeV$, since the entropy dilution factor from saxion decay is determined by the temperature ratio: $\Delta = T_{e}^{s} / T_{D}^{s}$~\cite{Kolb:1990vq}.
The decay temperature of $T_{D}^{s} \simeq 53 \GeV$ is realized when saxion with $m_{s} = 110 \GeV$ decays dominantly into $b {\bar b}$, and $v_{\rm PQ} = 2.5 \times 10^{10} \GeV$~\cite{Bae:2013hma}.
In this case, the total axino density is dominated by the freeze-in contribution.
Consequently, we find that the total mass density of $7 \keV$ axinos also meets {\it the observed DM one}, {\it i.e.},
\begin{equation}
\Omega_{\axino}h^{2} \simeq 0.1
\left( \frac{4.7}{\Delta} \right) \left( \frac{2.5 \times10^{10} \GeV}{v_{\rm PQ}} \right)
\left( \frac{m_{\axino}}{7 \keV} \right) \,.
\end{equation}

{\it Conclusions} -- We have examined $7 \keV$ axino DM in the RPV SUSY DFSZ model by incorporating the 3.5 keV X-ray line excess and the Ly-$\alpha$ forest constraints.
The model naturally introduces two key ingredients:
1) the $\mu$-term interaction, which is responsible for freeze-in production of axinos and 2) the bRPV term, which is responsible for axino-neutrino mixings.
While the $3.5 \keV$ line excess is easily explained by $7 \keV$ axino DM decay via an axino-neutrino mixing, the constraints from the Ly-$\alpha$ forest data impose a colder phase space distribution on axino DM.
Freeze-in production of axinos via the $\mu$-term interaction indeed leads to a colder phase space distribution.
As a result, the axino phase space distribution meets the most stringent limit from the Ly-$\alpha$ forest data with the mild entropy production from late-time saxion decay, which is inherent in the model.
We stress that, even with entropy production, the whole DM density is explained and dominated by the freeze-in axinos.

The result shown in this letter implies that X-ray observations determine the axino mass and its mixing parameter with active neutrinos, while Ly-$\alpha$ forest data and the observed DM mass density narrow down the saxion mass as well as the PQ breaking scale. 
Once the observational aspects of freeze-in axinos become evident, we can constrain and probe the underlying PQ breaking sector and its communication with the SUSY breaking sector.
We also emphasize that our analysis of the resultant axino phase space distribution and linear matter power spectrum can be easily applied to other freeze-in DM models.

{\it Acknowledgements} -- The work of KJB and AK was supported by IBS under the project code IBS-R018-D1.
SPL has received support from the Marie-Curie program and the European Research Council and Horizon 2020 Grant, contract No. 675440 (European Union).
AK and SPL would like to acknowledge the Mainz institute for Theoretical Physics (MITP) where this work was initiated.

%\bibliographystyle{utphys}
%\bibliography{wdm_letter}

\end{document}